\documentstyle[epsf,twocolumn]{jpsj}
\author{Yoshiki {\sc Imai} and Norio {\sc Kawakami}}
\title{
Correlation Effects on the Double Exchange Model
in a Ferromagnetic Metallic Phase}
\inst{Department of Applied Physics,
Osaka University, Suita, Osaka 565-0871}

\recdate{}

\abst{
The effect of the Hubbard interaction among conduction
electrons on the double exchange model is investigated
in a ferromagnetic metallic phase.
Applying iterative perturbation theory to the Hubbard interaction
within dynamical mean field theory,
we calculate the one-particle spectral function and the optical
conductivity,
in which  coherent-potential approximation
is further used to treat the ferromagnetic Hund coupling between
conduction electrons and localized spins.
Identifying the decrease of the magnetization for the localized spin
with the increase of the temperature,
we discuss the temperature dependence of the one-particle
spectrum and the optical conductivity. It is found that the interplay
between the Hund coupling and the Hubbard interaction
dramatically changes  the spectral function,
while it is somehow obscured in the optical conductivity.
}

\kword{correlated electron systems, double exchange model,
iterative perturbation theory}

\def\runtitle{Correlation Effects on the Double Exchange Model}
\def\runauthor{Yoshiki {\sc Imai} and Norio {\sc Kawakami}}

\begin{document}
\sloppy
\maketitle

\section{Introduction}

There has been a resurgence of interest in doped perovskite manganese oxides
such as R$_{1-x}$A$_{x}$MnO$_{3}$
(R is rare-earth and A is alkaline-earth ions), since the discovery
of the colossal magnetoresistance.
These compounds show a variety of interesting phenomena,
\cite{Tokura-I,Tokura-II,Tokura-III,Tokura-IV,Tokura-V,Ramirez}
which are caused by several competing interactions including
not only spin but also charge and orbital degrees of freedom.
\cite{Zener,Anderson-Hasegawa,Furukawa,Shiba-I,Shiba-II,Ishihara-Nagaosa,Nak
ano,Motome,Horsch,Khaliullin-I,Ferrari,Held,Ishihara-Maekawa,Maezono-I}
Since the ferromagnetic state
in a metallic phase is stabilized by the Hund coupling
between localized spins and conduction electrons,
the double exchange model has been used as
a proper model to describe the ferromagnetic metallic manganites.
\cite{Zener,Anderson-Hasegawa,Furukawa}
Various unusual properties observed
experimentally have been clarified by using this model, in which
the Coulomb interaction among conduction electrons has been
neglected for simplicity\cite{Furukawa,Shiba-I,Shiba-II}, or
has been assumed to be infinitely strong.
\cite{Ishihara-Nagaosa}
However, the finite Coulomb interaction should play an important
role for the dynamics in  Mn oxides, stimulating
further  theoretical studies to take into account
 the finite  Hubbard interaction $U$, which have been
done for  the fully polarized case.
\cite{Nakano,Motome,Horsch,Khaliullin-I,Ferrari}
More recently, Held and Vollhardt have
studied the effect of the finite Hubbard interaction on a
paramagnetic metallic phase by means of dynamical mean field theory
(DMFT) and have claimed that such correlation effects indeed play
an important role together with the Hund coupling.\cite{Held}

In this paper, we study electron correlation effects on
the double exchange model in a ferromagnetic metallic phase
by introducing the finite Hubbard interaction among conduction electrons.
The present study is based on the approach of
Shiba et al.\cite{Shiba-I,Shiba-II} with coherent potential
approximation (CPA), and is in some respects complementary to the work
done by Held and Vollhardt for a paramagnetic phase.\cite{Held}
By exploiting iterative perturbation theory (IPT)
within DMFT for the  Hubbard interaction, we
compute the one-particle spectrum and the optical conductivity
in which we further use CPA to deal with
the Hund coupling. Based on the obtained results,
we qualitatively discuss the temperature dependence of
the above-mentioned dynamical quantities.
It is shown that the introduction of the Hubbard interaction
 dramatically changes the
one particle spectral function in the double exchange model.
We also find that the optical conductivity is affected by the
interplay of these two interactions, although it is not so
conspicuous as in the one-particle spectrum.

\section{Iterative Perturbation Theory Approach}

We study the double exchange model with the on-site Hubbard interaction
in three dimensions.  The Hamiltonian is given by
\begin{eqnarray}
H=\sum_{\langle i,j \rangle \sigma}t_{ij}c_{i\sigma}^{\dagger}c_{j\sigma}
-J\sum_{i}{\mib \sigma}_{i}\cdot {\mib S}_{i}
+U\sum_{i}n_{i\uparrow}n_{i\downarrow},
\end{eqnarray}
where $c_{i\sigma}^{\dagger}(c_{i\sigma})$ is the creation (annihilation)
operator of a conduction electron with the spin $\sigma
(=\uparrow,\downarrow)$
at the site $i$, and $n_{i\sigma}=c_{i\sigma}^{\dagger}c_{i\sigma}$.
For simplicity, the orbital degeneracy is neglected
in this paper. The Hund coupling $J$ between a localized
spin ${\mib S}_{i}$ ($S=1)$
and a conduction electron (${\mib \sigma}_{i}$ is the Pauli matrix)
is assumed to be ferromagnetic.

We shall treat the electron correlations due to the
Hubbard interaction within DMFT\cite{DMFT-Rev},
which is known as a powerful method to systematically study
strongly correlated electron systems. The treatment with DMFT
becomes exact in the limit of large spatial dimensions
\cite{Metzner&Vollhartd,Muller-I},
and even for three dimensions, it provides a powerful method which
has successfully described  a number of important phenomena,
such as the Mott transition, in strongly correlated electron systems.
\cite{Mott(Kotliar),Mott(Jarrel-I),Jarrel}
To perform the DMFT procedure, we further make use of
IPT (see below),\cite{Mott(Kotliar),IPT1,IPT3,arb-fill,Saso}  which
correctly reproduces  the self-energy both in the weak coupling limit as
well as the atomic limit, and furthermore properly interpolates the two
limiting cases.

We use CPA to treat the ferromagnetic Hund coupling.
The validity of applying CPA to the double exchange model
 is summarized as follows\cite{Shiba-II}.  We are now dealing with
 the ferromagnetic Hund coupling, so that
the dynamics of localized spins may be described rather well
by that of classical spins, because the ferromagnetic order considerably
suppresses quantum effects.
It may be thus legitimate to replace
the local spin ${\mib S}_{i}$ by the classical Ising spin, and
to further use CPA for treating Hund coupling for partially
polarized ferromagnetic phase,\cite{Shiba-II} in which
random potentials are induced by thermal fluctuations
of the Ising spins.
Note that  the CPA procedure is properly embedded in the
self-consistent equations  for DMFT \cite{CPA+IPT},
because both methods are based on local approximations.

We carry out the following self-consistent procedure for IPT combined
with CPA. Since the original lattice problem is converted to an
effective single impurity problem in DMFT,
let us first introduce the cavity Green function
${\cal{G}}_{\sigma}(\omega)$  and the
dressed Green function $G_{\sigma}(\omega)$,
\begin{eqnarray}
&& {\cal{G}}_{\sigma}(\omega)=
[\omega+{\rm i}\delta+\mu-\tilde{E}_{f\sigma}-\Delta_{\sigma}(\omega)]^{-1},
\\
&&G_{\sigma}(\omega)=
[{\cal{G}}_{\sigma}^{-1}(\omega)-\Sigma_{U\sigma}(\omega)]^{-1},
\label{eqn:dress-G}
\end{eqnarray}
where $\mu$ and $\Delta_{\sigma}(\omega)$ are the  chemical potential and
the renormalized hybridization function, respectively. Here we have denoted
the self energy due to the Hubbard interaction as $\Sigma_{U\sigma}$
and $\tilde{E}_{f\sigma}=E_{f}+Un_{\bar{\sigma}}$ where
$E_{f}$ is the impurity level in the reduced single-site problem.
Using the cavity Green function,
the second-order self-energy due to the Hubbard interaction
is expressed in a usual form,
\begin{eqnarray}
&\Sigma_{U\sigma}^{(2)}&(\omega)=U^{2}
\int_{-\infty}^{\infty}{\rm d}\epsilon_{1}
\int_{-\infty}^{\infty}{\rm d}\epsilon_{2}
\int_{-\infty}^{\infty}{\rm d}\epsilon_{3}\nonumber \\
&\times&
\rho_{\sigma}(\epsilon_{1})
\rho_{\bar{\sigma}}(\epsilon_{2})
\rho_{\sigma}(\epsilon_{3})\nonumber \\
&\times&
\frac{
f(\epsilon_{1})f(-\epsilon_{2})f(\epsilon_{3})+
f(-\epsilon_{1})f(\epsilon_{2})f(-\epsilon_{3})}
{\omega+{\rm i}\delta-\epsilon_{1}+\epsilon_{2}-\epsilon_{3}}
\end{eqnarray}
with $\rho_{\sigma}(\epsilon)
=-1/\pi{\rm Im}({\cal{G}}_{\sigma}(\epsilon))$, where
$f(\epsilon)$ is the  Fermi distribution function.
For arbitrary filling,
we improve the approximation by modifying the
second-order self energy as,
\begin{eqnarray}
\Sigma_{U\sigma}(\omega)=
\frac{A_{\sigma}\Sigma_{U\sigma}^{(2)}(\omega)}
{1-B_{\sigma}\Sigma_{U\sigma}^{(2)}(\omega)}
\end{eqnarray}
with
\begin{eqnarray}
A_{\sigma}=\frac{n_{\bar{\sigma}}(1-n_{\bar{\sigma}})}
{n_{0\bar{\sigma}}(1-n_{0\bar{\sigma}})}\,,\,
B_{\sigma}=\frac{(1-n_{\bar{\sigma}})U+E_{f}+\mu}
{n_{0\bar{\sigma}}(1-n_{0\bar{\sigma}})U^{2}}
\end{eqnarray}
where $n_{0\sigma}$ and $n_{\sigma}$ are the particle numbers
respectively defined  by
$n_{0\sigma}=-1/\pi\int_{-\infty}^{\infty}{\rm d}\omega
f(\omega){\rm Im}{\cal{G}}_{\sigma}(\omega)$ and
$n_{\sigma}=-1/\pi\int_{-\infty}^{\infty}{\rm d}\omega
f(\omega){\rm Im}G(\omega)$.
\cite{arb-fill,Saso}
Note that the coefficients $A_{\sigma}$ and $B_{\sigma}$ have been
introduced so as  to reproduce the correct results in
 the high energy limit and the atomic limit, respectively.

We now incorporate the effect of the Hund coupling by
introducing an additional self energy $\Sigma_{H\sigma}$, which
enters in the local Green function as,
\begin{eqnarray}
&&\widetilde{G}_{\sigma}^{\rm loc}(\omega)\nonumber \\
&=&\frac{1}{N}\sum_{k}
\widetilde{G}_{\sigma}(k,\omega)\nonumber \\
&=&\int_{-\infty}^{\infty}{\rm d}\epsilon
\frac{N_{0}(\epsilon)}
{\omega+{\rm i}\delta-\epsilon+\mu-\tilde{E}_{f\sigma}
-\Sigma_{U\sigma}(\omega)-\Sigma_{H\sigma}(\omega)},\nonumber \\
\label{eqn:IPT-local-G}
\end{eqnarray}
where $\widetilde{G}_{\sigma}(k,\omega)$ denotes
the one-particle Green function in the lattice system.
We employ the semielliptic form
as the bare density of states,
$N_{0}(\epsilon)=2/(\pi D^{2})\sqrt{D^{2}-\epsilon^{2}}$
with the bandwidth $D$.
Following a standard CPA procedure,
the average of the random scattering of conduction electrons
due to localized spins should be zero, leading to
\begin{eqnarray}
{\Big \langle}
\frac{-JS_{i}\sigma -\Sigma_{H\sigma}(\omega)}
{1-\widetilde{G}_{\sigma}^{\rm loc}(\omega)
(-JS_{i}\sigma -\Sigma_{H\sigma}(\omega))}
{\Big \rangle}_{S_{i}}=0,
\label{eqn:average2}
\end{eqnarray}
which is rewritten in terms of the magnetization $M$ of localized
spins as,
\begin{eqnarray}
&\widetilde{G}_{\sigma}^{\rm loc}&(\omega)\Sigma_{H\sigma}(\omega)+
\Sigma_{H\sigma}(\omega)\nonumber \\
&-&\widetilde{G}_{\sigma}^{\rm loc}(\omega)(JS)^{2}
+\sigma JM=0.
\end{eqnarray}
The self-consistency condition requires the
local Green function in Eq. (\ref{eqn:IPT-local-G}) to be equivalent to
 the dressed impurity Green function in Eq. (\ref{eqn:dress-G}).
We thus obtain,
\begin{eqnarray}
\Delta_{\sigma}(\omega)
=\frac{D^{2}}{4}\widetilde{G}_{\sigma}^{\rm loc}(\omega)
+\Sigma_{H\sigma}(\omega).
\end{eqnarray}
Note that the chemical potential $\mu$ should be determined
 by the Friedel sum rule
in the reduced impurity problem.  This completes the description of
our IPT approach combined with CPA.

\section{Correlation Effects on Dynamical Quantities}

\begin{figure}[h]
\epsfxsize=9cm
\centerline{\epsfbox{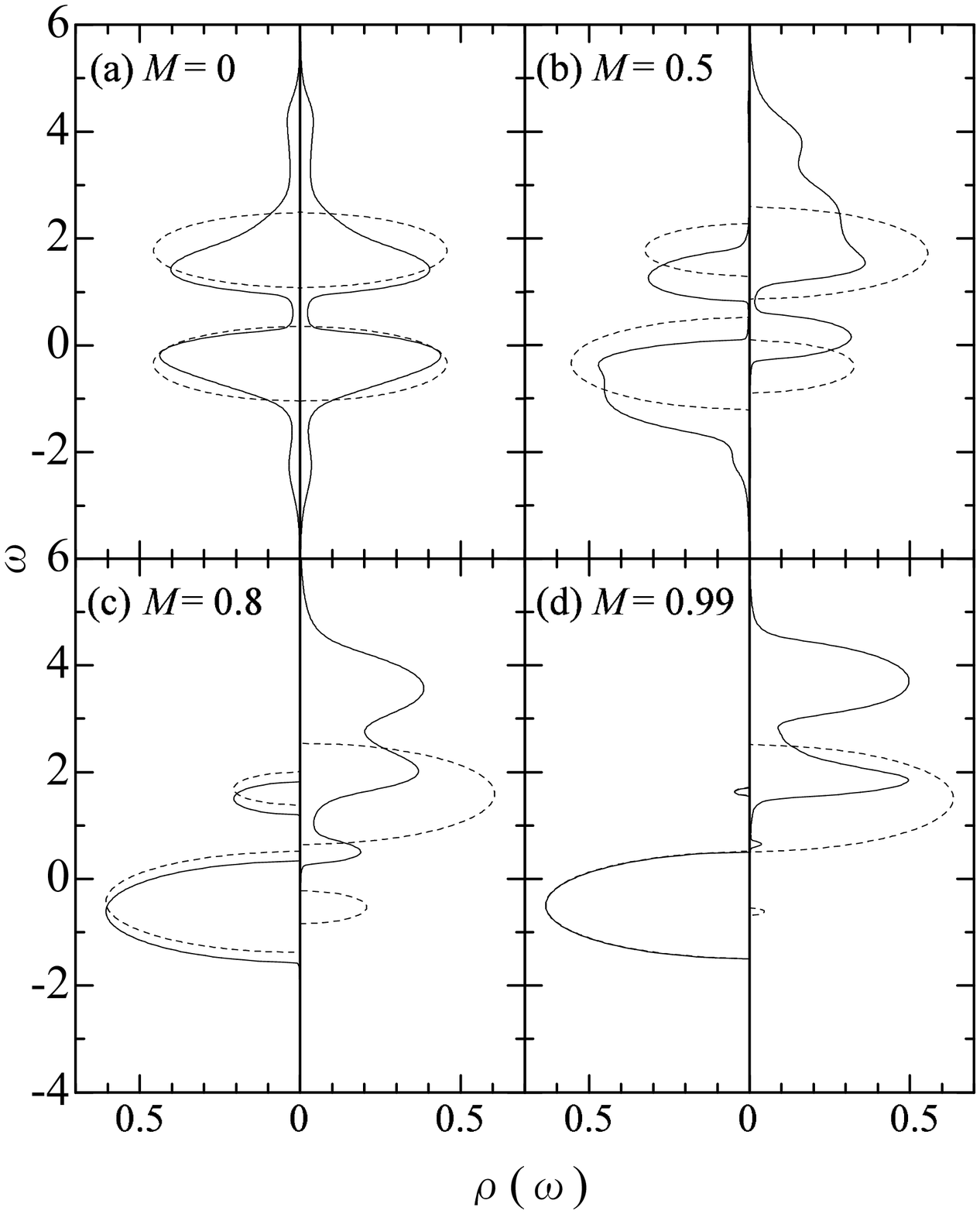}}
\vspace{-25mm}
\caption{One-particle spectral function $\rho(\omega)$ calculated by
IPT and CPA;
(a) $M=0$ (Curie temperature), (b) $M=0.5$,
(c) $M=0.8$, (d) $M=0.99$ ($T\sim$ 0K).
The hole (electron) concentration is $x=0.2$ (0.8).
In all figures, the left (right) panel represents the spectrum
for up (down) spin electrons.
The energy is measured from the  Fermi level $\omega=0$, and
the bandwidth $D$ is taken to be unity.
The Hund coupling is taken as $J=1$, and
the Hubbard interaction is set to be $U=2$ (solid line)
and $U=0$ (dashed line).
}
\label{fig:DOS}
\end{figure}

We have performed the numerical calculation by
iterating the procedure mentioned in the previous section
until each quantity should converge within a desired accuracy.
The one-particle spectrum of conduction electrons thus
computed is shown in Fig. \ref{fig:DOS}. We first mention
that since we are dealing with the dynamics of localized spins
by CPA, the change in the magnetization may be
approximately considered to be caused by thermal
fluctuations.\cite{Shiba-II}   Namely,
the magnetization of localized spins $M$ is
increased from (a) to (d), which implies
in our treatment that the temperature is monotonically decreased
from $T=T_{C}$ (critical temperature) down to $T=0$.

\begin{figure}[h]
\epsfxsize=9cm
\centerline{\epsfbox{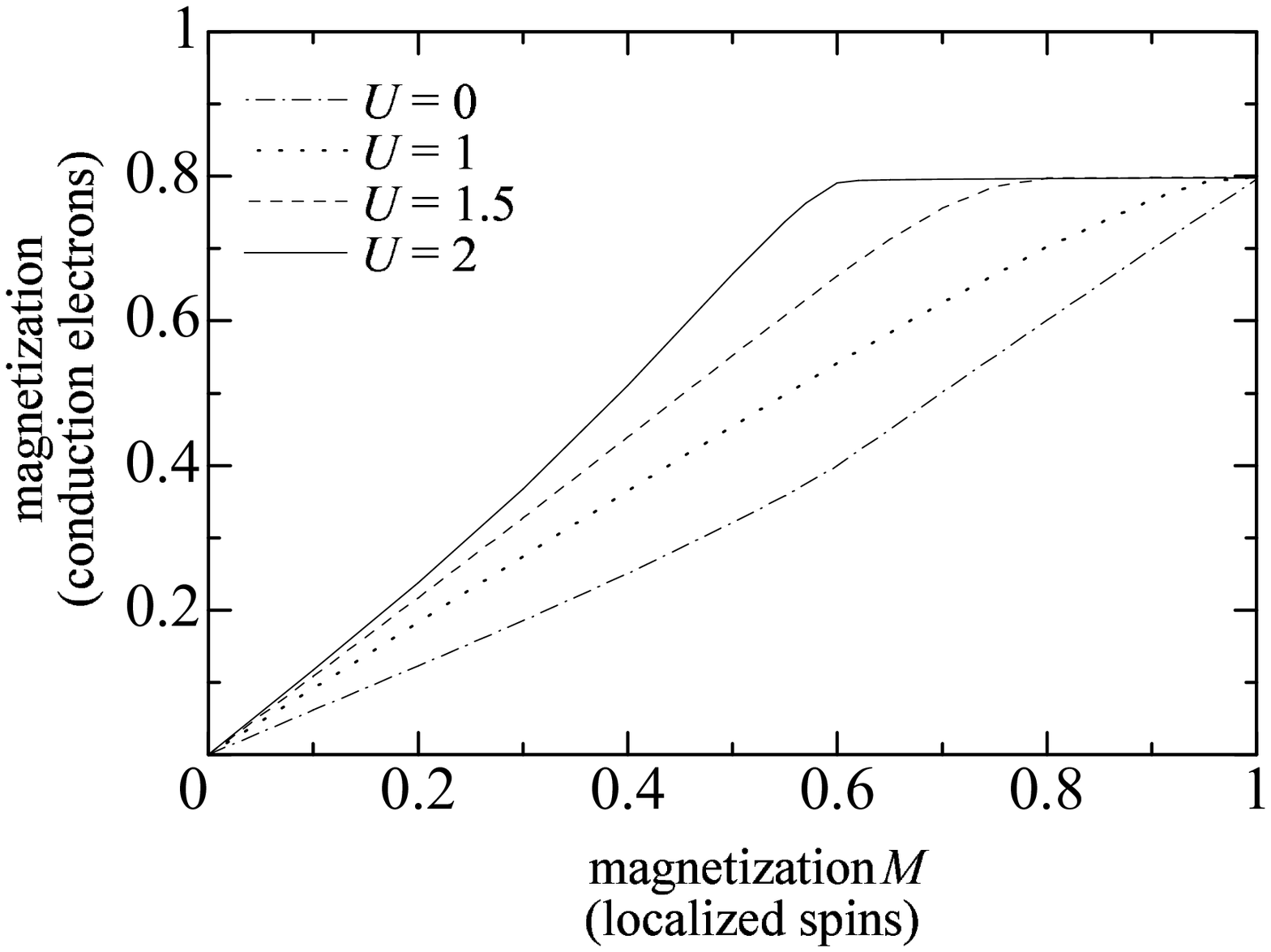}}
\vspace{-65mm}
\caption{Plot of the magnetization for conduction electrons as a function
of that for localized spins $M$ for various $U$.
The hole concentration is $x=0.2$.
The other parameters are as in Fig. \ref{fig:DOS}.
}
\label{fig:mag}
\end{figure}

We start our discussions with the case of (a), where there is no
magnetization in the system. In this paramagnetic phase,
the spectrum near the Fermi surface in the $U=2$ case
gets narrower than the non-interacting case.
In addition to this band-narrowing effect, there is
the life-time effect, which is accompanied by a small
structure developed around the energy for the doubly occupied
state. These aspects are in agreement with those obtained by Held and
Vollhardt.\cite{Held} With the increase of $M$ (decrease
of the temperature),  the spectral function
 shows quite different behaviors for up- and down-spin electrons.
The spectral function for up-spin electrons is gradually
reduced to the original semielliptic form,  whereas
that for down-spin electrons starts to develop several hump structures,
whose origin will be discussed below momentarily.
Note also in (b) that the spectrum with up- (down-) spin electrons in
the $U=2$ case shifts to a lower (higher) energy side compared
with the noninteracting case, which is  caused by the interplay
between the Hubbard and the Hund interactions. Namely,
since the magnetic susceptibility of conduction electrons is enhanced
by the Hubbard interaction, the magnetization
of conduction electrons for a given $M$ of localized spins
becomes larger for larger $U$ via the Hund coupling, giving rise to
the large self-energy shift observed in (b).  We will see below that
this shift affects the profile of the optical conductivity.
For reference, we show the magnetization of conduction electrons
as a function of $M$
for several choices of $U$ in Fig. \ref{fig:mag}.

\begin{figure}[h]
\epsfxsize=9cm
\centerline{\epsfbox{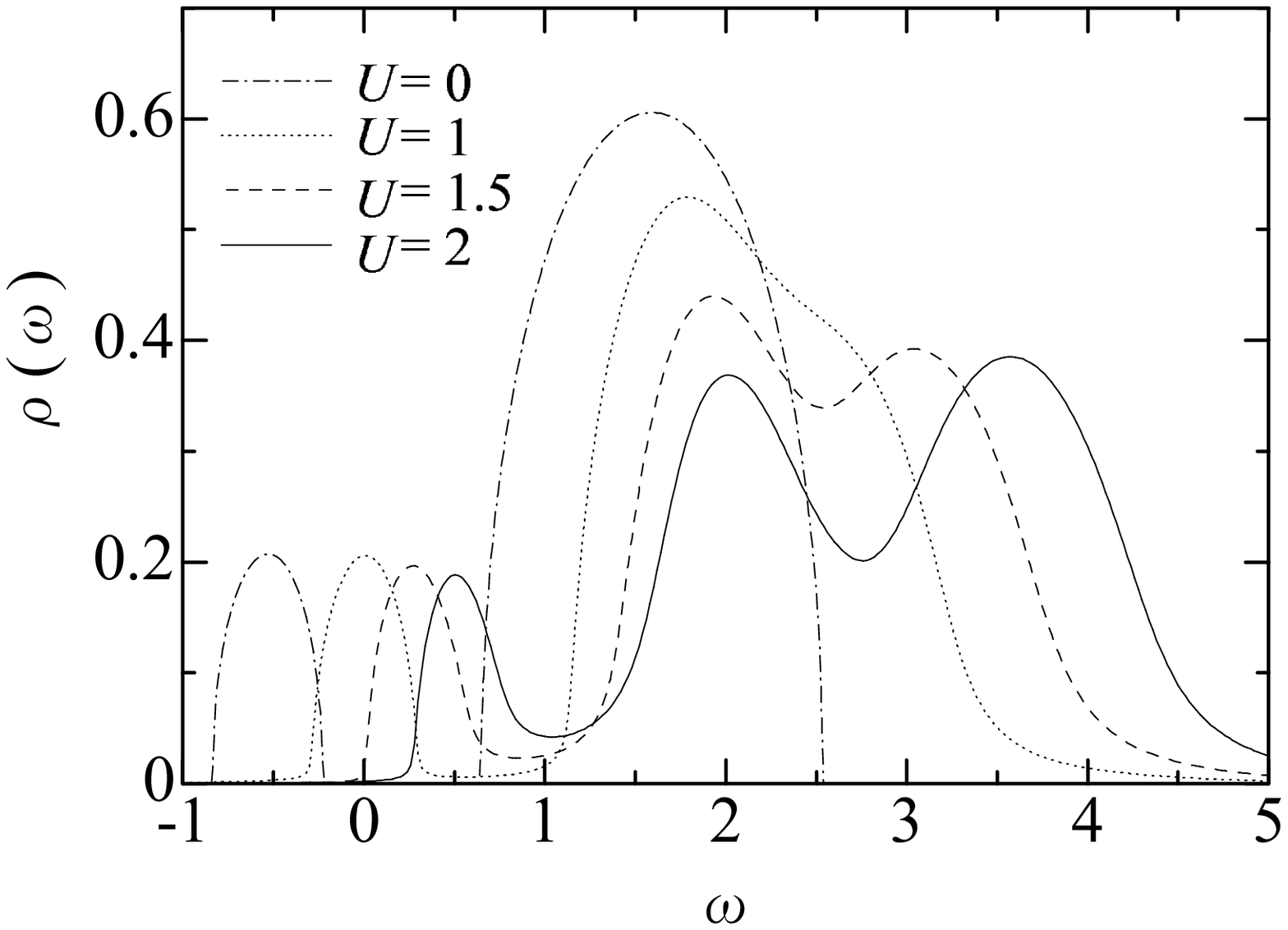}}
\vspace{-70mm}
\caption{One-particle spectral function $\rho(\omega)$
for down spin electrons for various $U$ in the case of $M=0.8$.
The other parameters are as in Fig. \ref{fig:DOS}.
}
\label{fig:M=0.8}
\end{figure}

As $M$ is further increased, the spectral function for down-spin electrons
develops three distinct peaks (Fig. \ref{fig:DOS}(c)). To see the origin
of the structure, we first recall that conduction electrons in (c)
are almost fully polarized, though localized spins are still
partially polarized ($M=0.8$), as a consequence of the enhanced spin
susceptibility (see Fig. \ref{fig:mag}).
Thus adding a down-spin electron
 to an unoccupied site may have two possibilities depending on whether
the corresponding localized spin is parallel or antiparallel
to the spin of the added electron. The former contributes to
the lowest-energy peak where the added electron gains the energy
$-J$, while  the latter forms the middle peak
with the energy increment $J$ for making antiparallel spins.
On the other hand, if a down-spin
electron is added to the site already occupied by an up-spin electron,
this makes the highest-energy peak in the spectrum.
Note that the position of the lowest-energy peak is pushed from the
bare level up to slightly above the Fermi level, which is
again caused by the self-energy shift due to the Hubbard interaction and
the Hund-coupling.
To clearly show how the above three-peak structure is
developed with the increase of  the Hubbard interaction $U$,
we display of $M=0.8$ for several choices of $U$ in Fig. \ref{fig:M=0.8}.
It is seen that the above-mentioned three-peak structure is gradually formed
with the increase of $U$, which is accompanied by the rather large
self energy shift.
Finally, in the case of (d), the localized spins are almost fully polarized,
so that the spectrum for up-spin electrons is reduced to
the non-interacting one,
while the spectrum for down spin electrons still has the
two-peak structure.  As mentioned above,
the lower peak represents a particle-addition spectrum to
an empty site, while the upper peak corresponds to that
for a singly occupied site.
In all the cases (a)$\sim$(d),
the interplay between the Hund coupling and
the Hubbard interaction is important to determine the temperature
dependence in the profile of the one-particle spectral function.

Experimentally, several nontrivial effects on
the photoemission spectrum have been reported in the manganese
compounds,\cite{photo-I,photo-II,photo-III}
which may not be simply explained without the Coulomb interaction.
A typical example is that the band width
observed in a paramagnetic phase is
about twice as large as that expected from the band calculation.
This may be caused by the correlation effects
among conduction electrons,
as recently explained  by Held and Vollhardt
with the use of DMFT.\cite{Held}
Our results in Fig. \ref{fig:DOS}(a)
are  consistent with theirs as well as
the experimental tendency.  Also, it has been reported that
the coherent spectrum near the Fermi surface is increased as
 the temperature is decreased below
the critical temperature, indicating the decrease of effective
electron correlations at low temperatures.\cite{photo-I,photo-II,photo-III}
Such effects are indeed seen in the present results of  Fig. \ref{fig:DOS}
for electrons with up spin, namely, the decrease of the temperature
gradually reduces the effective interaction among conduction electrons in
the
filled band, being consistent with the experiments.
Although our treatment here neglects several important effects due to
the orbitally degenerate bands with anisotropic hoppings, etc,
it turns out that some qualitative features  obtained are
consistent with those observed in the photoemission spectrum.

\begin{figure}[h]
\epsfxsize=9cm
\centerline{\epsfbox{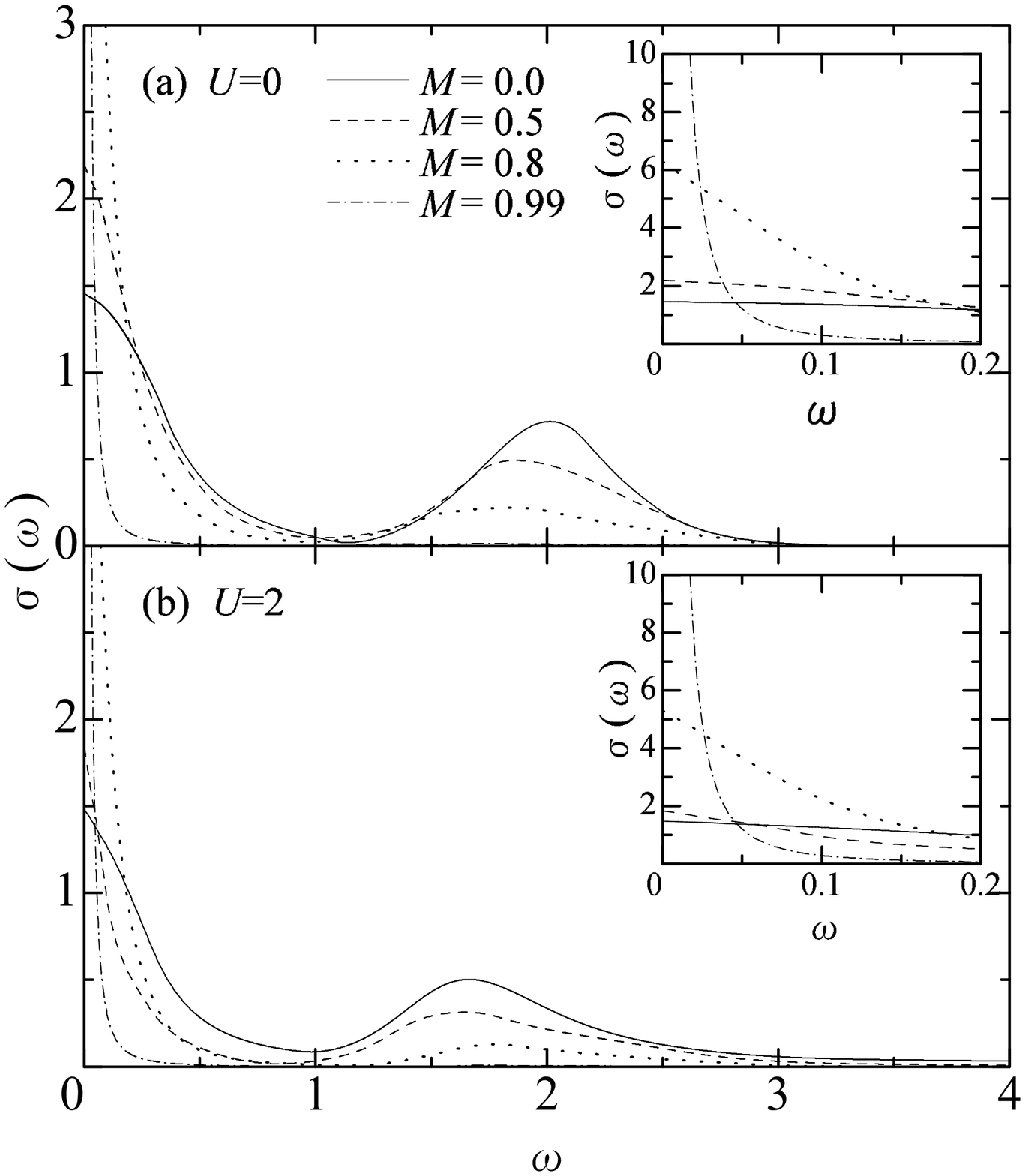}}
\vspace{-30mm}
\caption{Optical conductivity $\sigma(\omega)$
for (a) $U=0$, (b) $U=2$;
$M=0$ (solid), $M=0.5$ (dashed), $M=0.8$ (dotted)
and $M=0.99$ (dash-dotted), respectively.
The insets show the low-frequency part.
The other parameters are as in Fig. \ref{fig:DOS}.}
\vspace*{-0.5cm}
\label{fig:opt}
\end{figure}

In order to further investigate the dynamical response in the system,
we calculate the optical conductivity.
We employ the following formula for the optical conductivity
in which the vertex correction is neglected,
\begin{eqnarray}
&\sigma&(\omega)=\pi \sum_{\sigma}
\int_{-\infty}^{\infty} {\rm d}\epsilon'
\int_{-\infty}^{\infty} {\rm d}\epsilon \nonumber \\
&\times&N_{0}(\epsilon)
A_{\sigma}(\epsilon,\epsilon')
A_{\sigma}(\epsilon,\omega+\epsilon')
\frac{f(\epsilon')-f(\epsilon'+\omega)}{\omega},\\
&A_{\sigma}&(\epsilon,\omega)=-\frac{1}{\pi}\nonumber \\
&\times&{\rm Im}{\Big (}\frac{1}{
\omega+{\rm
i}\delta-\epsilon+\mu-\tilde{E}_{f\sigma}-\Sigma_{U\sigma}(\omega)
-\Sigma_{H\sigma}(\omega)}{\Big )}.\nonumber \\
\label{eqn:optical-conductivity}
\end{eqnarray}
Note that this expression becomes exact in the limit of large dimensions.
\cite{Muller-I,opt1,Khurana}
The computed results are shown in  Fig. \ref{fig:opt}.
As well known, the optical conductivity consists of  two peaks,
each of which corresponds to the Drude-like part
and the interband part for which two separated bands are
formed  by the Hund coupling.
Comparing the $U=2$ case with the non-interacting one, we can see
that the structure near the $\omega \sim 0$ (Drude-like peak) gets somewhat
narrower via the correlation effect due to the
Hubbard interaction.  Also, we find that the  hump structure
corresponding to the interband excitations are smeared by the
life-time effect due to the Hubbard interaction $U$, and
this effect becomes more conspicuous when $U$ is large, being
consistent with the results
deduced for a paramagnetic phase.\cite{Held}. It should be noticed
here that the maximum position of the interband excitations
in the  $U=2$ case shows a non-monotonic behavior with
the increase of the magnetization $M$,
which is not observed in the noninteractning case.
Namely, the maximum position in the  $U=2$ case
is once lowered with the increase of the magnetization ($M=0.5$),
and then increased again up to the original position ($M=0.8$).
This is a consequence of the self energy shift induced by
the Hubbard interaction and the Hund coupling:
As mentioned in Fig. \ref{fig:DOS}(b), the one-particle
spectrum for $U=2$ suffers from the large self-energy shift
 when $M$ is increased.  If $M$ is further increased,
the self energy shift becomes small again, as seen in
Fig. \ref{fig:DOS}(c), (d), giving rise to the non-monotonic
behavior of the interband excitations as a function of $M$.
If the Hubbard interaction is further increased, the spectral
weight of the interband excitations becomes small.
In this way, the interplay of the Hubbard interaction and the Hund
coupling appears in the optical conductivity, although it is somewhat
obscured in contrast to a dramatic change in the one-particle spectrum.

\section{Summary}
We have studied how the correlation effect among conduction electrons
affects dynamical properties of the double exchange model in
a ferromagnetic metallic phase.
In order to investigate dynamical quantities
in the whole energy region, we have exploited iterative perturbation
theory  combined with coherent potential approximation for
the Hund coupling. By approximately relating
the decrease of the magnetization of the localized spin
with  the increase of the temperature,
we have discussed the temperature dependence of the one-particle
spectrum and the optical conductivity.
It has been shown that  the interplay between
the Hund coupling  and the Hubbard interaction
dramatically affects the profile of the one-particle spectrum, which
results in a characteristic structure quite
different from the non-interacting case.
Some results have been found to
be qualitatively consistent with photoemission experiments.
On the other hand, the effect of the Hubbard interaction
is somewhat smeared in the optical conductivity, although it
indeed affects the profile for  both of the Drude-like intraband  part
as well as the interband part.

We have shown that there appears a non-monotonic temperature
dependence in the one-particle spectrum as well as the optical
conductivity in a ferromagnetic phase, which is caused by
the self-energy shift due to the Hubbard
interaction and the Hund coupling.
For example, such a  behavior emerges in the
interband excitations in the optical
conductivity in Fig. \ref{fig:opt}.
As already pointed out, however, the interband peak-structure is
smeared when the Hubbard interaction is increased, presumably
making it rather difficult to observe such a non-monotonic
behavior. It seems thus interesting  to check whether this
type of the correlation effect could be observed experimentally in the
one-particle density of states around the Fermi level.

In this paper, we have neglected the orbital degeneracy
for conduction electrons as well as the
corresponding off-diagonal components for the
hopping matrix.  Correlation effects including the orbital degrees
of freedom may be expected to provide further nontrivial properties.
To this end, it is necessary
to extend the present study to such orbitally degenerate cases
in a ferromagnetic phase.

\section*{Acknowledgements}
We would like to thank S. Kumada and N. Furukawa for valuable discussions.
The work is partly supported by a Grant-in-Aid from the Ministry of
Education, Science, Sports, and Culture.

\newpage

\end{document}